\documentclass[twocolumn,showpacs,preprintnumbers,amsmath,amssymb]{revtex4}

\usepackage{graphicx}
\usepackage{dcolumn}

\usepackage{mathptmx, courier, pifont}
\usepackage[scaled=0.92]{helvet}
\usepackage[T1]{fontenc}
\usepackage{textcomp}

\usepackage{bm}

\newcommand{\singlefig}[6]{%
\begin{figure} \vspace{#3}%
\includegraphics*[scale=#5]{#2}%
\caption{\label{fig:#1} #6}%
\vspace{#4}%
\end{figure}}

\newcommand{\quarterthin}{\kern 0.0417em}

\newcommand{\pratio}{\sigma}

\begin{document}


\title
{ Inhomogeneity, Dynamical Symmetry, and Complexity in
High-Temperature Superconductors: Reconciling a Universal Phase
Diagram with Rich Local Disorder }

\author
{Mike Guidry$^{(1)}$, Yang Sun$^{(2)}$, and Cheng-Li Wu$^{(3)}$ }
\affiliation{ $^{(1)}$Department of Physics and Astronomy,
University of Tennessee, Knoxville, Tennessee 37996--1200
\\ $^{(2)}$Department of Physics, Shanghai Jiao Tong University,
Shanghai 200240, People's Republic of China
\\ $^{(3)}$Physics Department, Chung Yuan Christian University,
Chung-Li, Taiwan 320, ROC
}

\date{\today}

\begin{abstract}

A model for high-temperature superconductors incorporating
antiferromagnetism, $d$-wave superconductivity, and no double
lattice-site occupancy can give energy surfaces exquisitely balanced
between antiferromagnetic and superconducting order for specific
ranges of doping and temperature.  The resulting  properties can
reconcile a universal cuprate phase diagram with rich inhomogeneity,
relate that inhomogeneity to pseudogaps, give a fundamental
rationale for giant proximity effects and other emergent behavior,
and provide an objective framework for separating essential from
peripheral in the superconducting mechanism.

\end{abstract}

\pacs{74.20.-z, 71.10.-w}

\maketitle

High-temperature superconductivity  was discovered more than two
decades ago \cite{highTc_discovery}, but its interpretation remains
controversial \cite{bonn06}. We believe that this lack of
theoretical consensus  may be attributed to two fundamental, and
related, issues:  (1)~Most models emphasize a limited aspect of the
(complex) problem; as a result, there are few solvable models that
capture a sufficient range of essential physics. (2)~The complex
behavior of these compounds obscures the superconducting mechanism
because, in the absence of solvable models incorporating a wide
enough range of physics, it is difficult to separate essential
features from secondary ones using only data.

For example, high-temperature superconductors exhibit a variety of
spatial inhomogeneities such as stripes or checkerboards,
particularly in the hole-underdoped region and near magnetic vortex
cores \cite{hayd04,tran04,hoff02,vers04,hana04}. The relationship of
this inhomogeneity to the unusual properties of these systems is
unsettled.  Does it lead to superconductivity (SC), does it oppose
SC, or is it a sideshow? Particularly vexing is that this rich
variety of inhomogeneity would at first glance seem to be
inconsistent with broad evidence for a cuprate phase diagram that is
rather universal, particularly for hole-doped compounds.  How does
one reconcile a seemingly universal phase diagram with a bewildering
array of inhomogeneity for individual compounds?

Dopant atoms play a dual role in high-temperature superconductors:
they support SC globally by enhancing charge carrier density, but
may suppress SC locally through atomic-scale disorder. McElroy et al
\cite{mcel05} found strong disorder in atomically-resolved scanning
tunneling microscope  images of the SC gap for Bi-2212. They
concluded that this disorder derives primarily from dopant
impurities and that charge variation between nanoregions is small,
implying that inhomogeneity is tied to impurities need not couple
strongly to charge.

We show here that such inhomogeneities are a generic consequence of
perturbations on the antiferromagnetic (AF) and SC correlations,
largely independent of specifics and not necessarily coupled to
charge variation.  Further, we show that these properties are
consistent with a global cuprate phase diagram, are tied intimately
to the nature of pseudogap states, and imply a linkage among
pseudogaps, inhomogeneity, and emergent behavior.  Thus we provide a
testable hypothesis for separating primary from derivative features
in the rich high-temperature superconductor data set.


The SU(4) model \cite{guid99,guid04,wu05,gui07a,gui07b,sun07} is a
fermion many-body theory that incorporates AF and SC order on an
equal footing, conserves spin and charge, and implies no double
occupancy on the lattice (Mott insulator ground state at half
filling).  The effective SU(4) Hamiltonian can be expressed as
\begin{equation}
H =H_0 -\tilde{G}_0\, [ (1-\pratio)D^{\dagger }D +
\pratio \vec{Q}\cdot \vec {Q} ]
+ g' \vec S\cdot \vec S,
\label{eq1}
\end{equation}
where $H_0$, $\tilde G_0$ and $g'$ are constants, $D^\dagger$
creates $d$-wave singlet pairs, $\vec{Q}$ is the staggered
magnetization, $\vec S$ is spin,
$ \pratio = \sigma(x) = \chi(x)/(\chi(x)+G_0(x)), $ and $ \tilde G_0
= \chi(x)+G_0(x), $ where $G_0(x)$ and $\chi(x)$ are effective SC
and AF coupling strengths.  Doping is characterized by a parameter
$
x=1-n/\Omega
$
for an $n$-electron system, with $\Omega$ the maximum number of
doped holes (or doped electrons for electron-doped compounds) that
can form coherent pairs, taking the normal state at half filling as
the vacuum.
Generally,
$
   P \simeq 0.25 x,
$ where $P$ is the standard hole-doping parameter, normalized to the
number of copper sites \cite{guid04}.

Three dynamical symmetry limits have exact solutions \cite{guid99}.
The SO(4) limit ($\pratio=1$) is a collective AF state, the SU(2)
limit ($\pratio=0$) is a collective SC state, and the SO(5) limit
($\pratio= 1/2$) is a critical dynamical symmetry interpolating
between the SC and AF limits \cite{guid99}. For other values of
$\pratio$ an approximate solution can be obtained using generalized
coherent states. We characterize AF in these solutions by the
staggered magnetization $Q \equiv \langle {Q}_z \rangle$, or by
$\beta$ defined through
\begin{equation}
Q
= 2\Omega \beta (n / (2\Omega) - \beta^2)^{1/2},
\label{betadef}
\end{equation}
and singlet $d$-wave pairing through $\Delta = \langle
D^{\dagger}D\rangle^{1/2}$ .


Total energy surfaces are obtained from the expectation value of
Eq.\ (\ref{eq1}) in coherent state approximation \cite{guid99}. To
relate them to data we use the variation of $\pratio$ with doping,
which was determined by fitting to cuprate data and given in Fig.\ 1
of Ref.\ \cite{wu05}a. Figure \ref{fig:esurfaces} shows SU(4) energy
surfaces as functions of AF order $\beta$, SC order parameter
$\Delta$, and doping $x$. These energy surfaces exhibit two
fundamental instabilities that may play a large role in the
properties of cuprate superconductors. The first is an instability
against condensing pairs when the system is doped away from
half-filling infinitesimally.  This instability is similar to the
Cooper instability for a normal superconductor, but generalized to a
doped Mott insulator, and accounts for the remarkably rapid
development of superconductivity with hole doping in the cuprates
\cite{gui07a}. In this paper we concentrate on the properties and
implications of a second fundamental instability that occurs in the
underdoped region.



\singlefig {esurfaces} {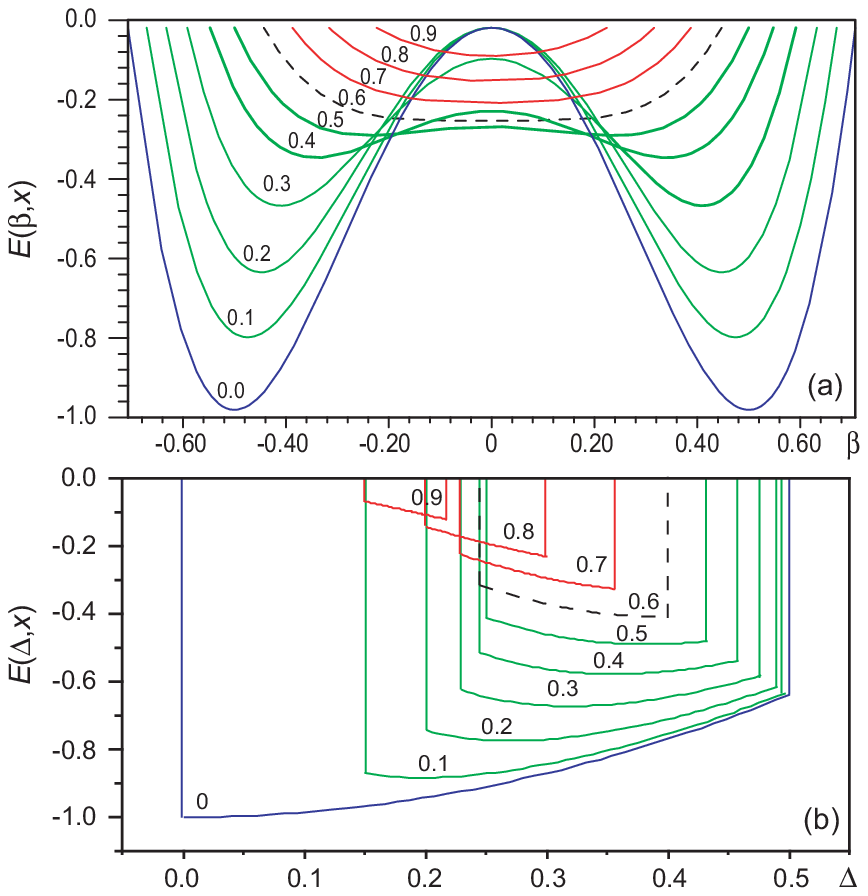} {0pt} {0pt} {0.89} {(Color
online) Total energy vs.\ (a) AF order $\beta$ and (b) SC order
$\Delta$; curves labeled by hole-doping $x\simeq 4P$, where $P$ is
the number of holes per copper site. Energy is in units of
$\chi\Omega^2/4$, with $\chi$ the AF coupling strength. The black
dashed line indicates the critical doping $x=x_q$ (see Ref.\
\cite{wu05}); color coding indicates energy surfaces favoring SC
(red), AF (blue), and AF + SC (green). The  doping-dependent ranges
of the energy contours in the order parameters reflect the finite
valence space (single band) of the model.}


The energy surfaces at constant doping fall into three general
classes: AF+SC (e.g., $x=0.1$), SC (e.g., $x=0.9$), and critical
(e.g., $x=0.6$, which marks a quantum phase transition).  Curves in
the AF+SC class have minima at finite and large $\beta_0$, and small
but finite $\Delta_0$, where the subscript zero denotes the value of
the order parameter at the minimum of the energy surface.  Curves in
the class SC are characterized by $\beta_0=0$ and finite $\Delta_0$.
Of most interest here are surfaces near critical in Fig.\
\ref{fig:esurfaces}, which correspond to broken SU(4) $\supset$
SO(5) dynamical symmetry \cite{guid99} and are  {\em flat} over
large regions of parameter space.  This implies that there are many
states lying near the ground state with very different values for
$\beta$ and $\Delta$.  Thus the surface is critically balanced
between AF and SC order, and small perturbations can drive it from
one to the other.  This defines a critical dynamical symmetry of the
SU(4) algebra \cite{guid99}; we shall term this situation {\em
dynamical criticality}.

The extreme sensitivity of critical surfaces to perturbations is
illustrated in Fig.\ \ref{fig:perturbations}.  Each set of curves is
associated with a fixed value of doping $x=0.6$ (equivalently,
$P=0.15)$, with the solid line corresponding to $\pratio=0.6$,  the
dashed line to a 10\% increase in $\pratio$ (AF perturbation), and
the dotted line to a 10\% reduction in $\pratio$ (SC perturbation).
The effect on the energy surface versus $\Delta$ (not shown) is
significant but less dramatic:  $\Delta_0$ is shifted, but remains
finite in all three cases.  We see that this small fluctuation in
$\pratio$ can alter the energy surface between AF+SC (finite
$\beta_0$ and $\Delta_0$) and SC ($\beta_0=0$ and finite
$\Delta_0$). This sensitivity is specific to the critical (broken
SU(4) $\supset$ SO(5)) dynamical symmetry.  The AF region near $x=0$
and the $d$-wave superconducting region at larger hole doping (see
Fig.\ \ref{fig:esurfaces}) are very stable against such
perturbations.

\singlefig {perturbations} {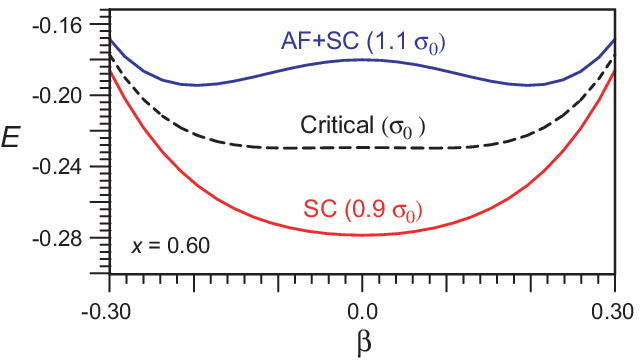} {0pt}
{0pt} {1.05} {(Color online) Energy surfaces as a function of the AF
order parameter $\beta$ for fixed doping $P = x/4 = 0.15$.  The
dashed line corresponds to $\pratio=\pratio_0 = 0.6$,  the upper
curve to a 10\% increase in $\pratio$, and the lower curve to a 10\%
reduction in $\pratio$. }

The AF instability displayed graphically in Fig.\
\ref{fig:perturbations} may also be understood analytically. From
the $T=0$ solution for $Q$ given by Eqs.\ (24a) and (14) of Ref.\
\cite{wu05}a, we find
\begin{equation}
    \left. \frac{\partial Q}{\partial x} \right|_{x=x_q} =
        \left. -\frac14 \frac{x_q + x_q^{-1}-2x}
        {[(x_q-x)(x_q^{-1} -x)]^{1/2}} \right|_{x=x_q}
        = - \infty ,
\label{analyticalDelderiv}
\end{equation}
and a small change in doping will cause a large change in
antiferromagnetic correlations near $x=x_q$.  This is a consequence
of SU(4) symmetry, which requires that $Q$ vanish for $x\ge x_q$ and
be finite for $0<x<x_q$.

It is not hard to conjecture mechanisms altering the ratio of AF to
SC coupling locally.  For example, Ref.\ \cite{mcel05} found that
nanoscale disorder is tied to influence of dopant impurities. Nunner
et al \cite{nunn05} (see also \cite{fang05}) compared these results
with Boboliubov--de Gennes calculations and proposed that
out-of-plane dopant atoms can modulate pairing on a scale comparable
to the lattice spacing, through lattice-distortion modification of
electron--phonon coupling or superexchange.


States associated with critical dynamical symmetry may be termed
{\em chameleon states:}  their variational energy surfaces are flat
over large regions of parameter space and their  intrinsic
collective properties may be changed qualitatively by a small
perturbing background that alters the AF--SC competition.  Figure
\ref{fig:esurfaces} suggests that underdoped cuprates have
near-critical energy surfaces. Thus, chameleon states are central to
the discussion of inhomogeneity and to the general issue of
understanding pseudogap states in underdoped cuprates.


Figure \ref{fig:stripeOrigin} is constructed from the expectation
value of (\ref{eq1}) in coherent state approximation, assuming a 1-D
spatial perturbation, $\sin(2\pi L)$, with $\sigma=0.6$.  It
illustrates schematically how a small (10\%) periodic fluctuation in
the AF and SC coupling for a critically symmetric underdoped
compound  can lead to inhomogeneity.  In this example, 1-dimensional
spatial variations of the coupling ratio $\pratio$ give fluctuations
in order parameters leading to stripes in which AF+SC
($\pratio>0.6$) and SC ($\pratio<0.6$) are favored alternately. Also
shown are the responses of  AF fluctuations $d\beta/dL$ to this
variation in $\pratio$. (We do not intend this as a realistic model
of a stripe phase, but as a cartoon indicating how such a model
could be built.)

\singlefig {stripeOrigin} {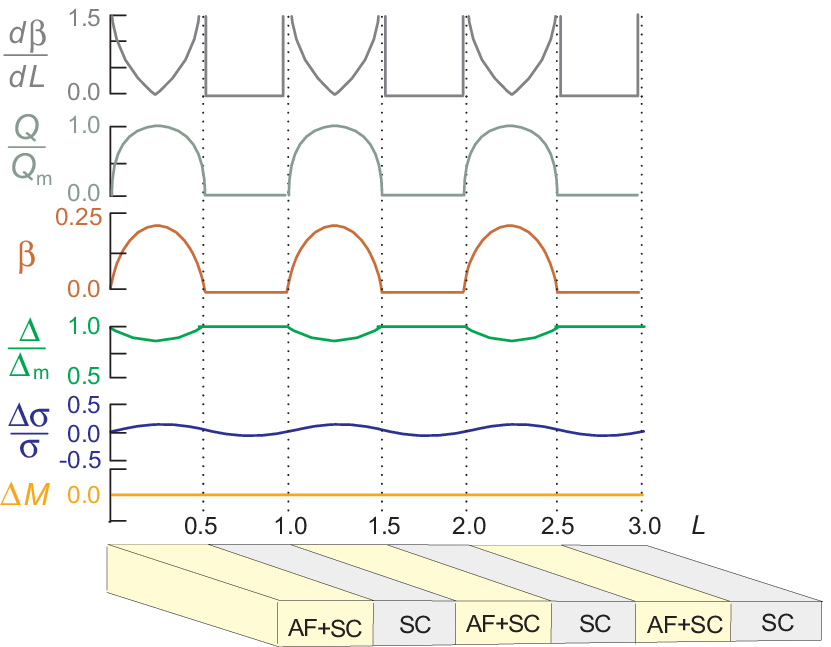} {0pt} {0pt}
{0.92} {(Color online) A small, spatially-periodic variation in the
AF and SC coupling can produce inhomogeneity without significant
charge variation. The charge is $M$ and a subscript m denotes
maximum values.}

Figure \ref{fig:stripeOrigin} indicates that a small spatial
variations in the coupling ratio $\pratio$ can produce regions
having large AF and weaker SC order, interspersed with regions
having significant SC correlations  but no AF order.  Although not
plotted, the AF+SC regions also exhibit small triplet pair
densities, which vanish in the SC regions.  The primary fluctuation
between stripes is in the AF order, which can jump between zero and
the maximum allowed by the SU(4) model between adjacent stripes; the
pairing order changes are much smaller and pairing is finite in both
the SC and SC+AF regions. The variation of $d\beta/d L$ indicates
that appreciable softness in AF and SC may occur on the boundaries
between regions.

The expectation value of the charge $M$ is not a function of the
coupling ratio $\pratio$ in the SU(4) coherent state, so the
critical energy surface fluctuations responsible for alternating
AF+SC and SC stripes in Fig.\ \ref{fig:stripeOrigin} cause no charge
variation ($\Delta M=0$). Data indicate that the relative charge
variation for Bi-2212 surface nanoscale patches is less than 10\%,
implying heterogeneity not strongly coupled to charge \cite{mcel05}.
This view is supported by analyses of heat capacity and NMR data on
Bi-2212 and YBCO that find a universal phase behavior for cuprates,
with little static charge modulation \cite{lora04, bobr02}.

Of course, the variation in $\pratio$ could itself be due to a
charge modulation. From the preceding discussion, we may expect that
if a charge modulation occurs in either the AF region near half
filling or the SC region at larger doping, its effect will be small
because the energy surfaces are not critical there.  However, if a
charge modulation occurs in the underdoped region where the energy
surfaces are near critical its effect could be amplified by
dynamical criticality even if $\pratio$ is not altered
significantly, since this is equivalent to a doping modulation.
Thus, we suggest a mechanism for producting strong inhomogeneity
without necessarily invoking charge fluctuations as the cause, but
that can in particular cases for underdoped compounds be produced by
a charge modulation. Such a mechanism could make understandable
strong inhomogeneity in the face of a universal phase diagram, and
resolve competing experimental claims regarding the role of charge
variation in producing inhomogeneity.

As an aside, inhomogeneity without charge modulation has an analog
in astronomy. We might assume, erroneously, that density in the
bright arms of a spiral galaxy (a form of ``stripe order'') is much
larger than that between the arms.  Actually, spiral arms are
prominent because the mass (gravitational charge) there is more
visible: galactic dynamics  generates many hot, luminous stars in
the spiral arms.  In analogy, the structure in Fig.\
\ref{fig:stripeOrigin} is not due to charge variation but rather to
modulation of the charge ``visibility'' by (quantum) dynamics.

The minimal patch size that can support SU(4) coherent states is
crucial to the present argument.  Experience  (e.g., Ref.\
\cite{FDSM}) suggests that dynamical symmetry can be realized in
fermion valence spaces having as few as several particles. This
would be consistent with inhomogeneities on scales comparable to
atomic dimensions, as required by data \cite{mcel05}.

Inhomogeneity caused by electronic self-organization is often
contrasted with that caused by dopant impurities.  Our discussion
implicates both as sources of nanoscale structure.  The proximate
cause may be impurities that perturb the SU(4) energy surface but
the criticality of that surface, which greatly amplifies the
influence of impurities, results from the self-organizing, doped
Mott insulator encoded in the SU(4) algebra.  Note a recent analysis
\cite{fang05} suggesting, from a different perspective, that
intrinsic amplification of impurity effects is required to explain
nanoscale structure in Bi-2212.


Critical dynamics may also produce inhomogeneities  near magnetic
vortices and magnetic impurities.  A magnetic field should suppress
SC relative to AF, so distance from a vortex $d$ may be expected to
alter the average value of $\pratio$, just as changing the doping
$P$ would. Therefore, we may expect a region near magnetic vortices
or impurities  where the symmetry is critical and exhibits
sensitivity to perturbations similar to that exhibited in Fig.\
\ref{fig:perturbations}, with $d$ modulating $\pratio$ rather than
$P$.


Giant proximity effects are observed in the cuprates where non-SC
copper oxide material sandwiched between superconducting material
can carry a supercurrent, even for a thickness much larger than the
coherence length \cite{bozo04}.  Phenomenology indicates that
pre-existing nanoscale SC patches can precipitate such effects
\cite{gonz05}.  We find similar possibilities on fundamental
grounds, but also suggest that the inhomogeneity need not pre-exist.
Dynamical criticality renders even a homogeneous pseudogap phase
unstable against large fluctuations in the AF and SC order. Thus,
proximity of superconducting material to pseudogap material coupled
with perturbations from background impurity fields can trigger
dynamical nucleation of nanoscale structure and giant proximity
effects, even if no static inhomogeneity exists beforehand.


In experiments with one unit cell thickness La$_2$CuO$_4$ AF barrier
layers between superconducting La$_{1.85}$Sr$_{0.15}$CuO$_4$
samples, Bozovic et al \cite{bozo03} found that the two phases did
not mix, with the barrier layer completely blocking a supercurrent.
These results were interpreted  to rule out many models of
high-temperature superconductors, in particular models in which SC
and AF phases are nearly degenerate like the SO(5) model
\cite{deml98}. The absence of a proximity effect between the AF and
SC phases (but presence of a strong proximity effect between
pseudogap and SC material) is plausibly consistent with the SU(4)
model: the AF phase is not rotated directly into the SC phase but
rather evolves with increased doping into a mixed SC and AF phase,
which then is transformed into a pure SC state at a critical doping
point near optimal doping, where the AF correlations vanish
identically \cite{wu05}.


The spontaneous appearance of properties that do not pre-exist in a
system's elementary components is termed emergence;  systems with
emergent behavior are said to exhibit complexity (see Dagotto
\cite{dag05} for a review).  Complexity can occur when the choice
between potential ground states is sensitive to even weak external
perturbations.
The amplification effect implied by SU(4) dynamical criticality can
facilitate emergent behavior and complexity.  The giant proximity
effect and the perturbatively-induced structure of Fig.\
\ref{fig:stripeOrigin} are examples. More generally, critical
dynamical symmetry may represent a fundamental organizing principle
for complexity in strongly-correlated fermi systems. For example,
critical dynamical symmetries are known in nuclear physics
\cite{zhan87}.


The SU(4) coherent state method that yields the variational energy
surfaces discussed here admits quasiparticle solutions that
generalize the BCS equations, giving a rich, highly universal phase
diagram in agreement with much available data \cite{wu05}, and a
quantitative description of fermi arcs \cite{gui07b}.  Therefore,
the propensity of the pseudogap state to a broad variety of induced
inhomogeneity, and a quantitative model of the cuprate phase diagram
(including pseudogaps) that exhibits highly universal character,
both follow directly from a model that implements AF and $d$-wave SC
competition in a doped Mott insulator.  This natural coexistence of
a universal phase diagram with a rich susceptibility to disorder
could reconcile many seemingly disparate observations in the cuprate
superconductors.  For example, since the SU(4) pseudogap state has
both pairing fluctuations and critical dynamical symmetry, it could
support Nernst vortex states as perturbations on a homogeneous phase
having incipient nanoscale heterogeneity.



In summary, competing antiferromagnetism and $d$-wave pairing,
constrained by no double site occupancy, can lead to energy surfaces
in hole-underdoped cuprates that are critically balanced between
antiferromagnetism and superconductivity. These surfaces can be
flipped between dominance of one or the other by small changes in
the ratio of the antiferromagnetic to pairing strength. Therefore
weak perturbations in the underdoped region, and near vortex cores
or magnetic impurities, can produce amplified inhomogeneity having
the spatial dependence of the perturbation but the intrinsic
character of an SU(4) symmetry. (The symmetry defines the possible
states; the perturbation selects among them.) Our results show that
such effects could, but need not, imply spatial modulation of
charge. More generally, we have suggested that critical dynamical
symmetry may be a fundamental principle of emergent behavior in
correlated fermion systems.

The generality of our solution implies that any realistic theory
describing superconductivity competing with antiferromagnetism
should contain features similar to those discussed here.  In
particular,  similar phenomena may be possible in the new iron-based
high-temperature superconductors \cite{sun08}.  Then the existence
of complex inhomogeneities for compounds having (paradoxically)
universal phase diagrams suggests that (1)~properties of
superconductors in the pseudogap state, and near magnetic vortices
and impurities, are largely determined by critical dynamical
symmetry, and (2)~inhomogeneity is a strong diagnostic for the
mechanism of high-temperature superconductivity but it is only
consequence, not cause.


We thank Takeshi Egami, Elbio Dagotto, Adriana Moreo, Pengcheng Dai,
and Boris Fine for discussions.


\bibliographystyle{unsrt}

\end{document}